\documentclass[jog]{igs}

\usepackage[utf8]{inputenc}

\usepackage{graphicx}
\usepackage{siunitx}
\usepackage{subcaption}

\usepackage{hyperref}
\usepackage{doi}

\begin{document}

\title[~]{{\it In situ}, broadband measurement of the radio frequency attenuation length at Summit Station, Greenland}

\author[Aguilar and others]{
J.~A.~Aguilar$^{1}$,
P.~Allison$^{2}$,
J.~J.~Beatty$^{2}$,
D.~Besson$^{3,4}$,
A.~Bishop$^{5}$,
O.~Botner$^{6}$,
S.~Bouma$^{7}$,
S.~Buitink$^{8}$,
M.~Cataldo$^{7}$,
B.~A.~Clark$^{9}$,
Z.~Curtis-Ginsberg$^{10}$,
A.~Connolly$^{2}$,
P.~Dasgupta$^{1}$,
S.~de Kockere$^{11}$,
K.~D.~de Vries$^{11}$,
C.~Deaconu$^{10}$,
M.~A.~DuVernois$^{5}$,
C.~Glaser$^{6}$,
A.~Hallgren$^{6}$,
S.~Hallmann$^{12}$,
J.~C.~Hanson$^{23}$,
B.~Hendricks$^{14}$,
C.~Hornhuber$^{3}$,
K.~Hughes$^{10}$,
A.~Karle$^{5}$,
J.~L.~Kelley$^{5}$,
I.~Kravchenko$^{15}$,
R.~Krebs$^{14}$,
R.~Lahmann$^{7}$,
U.~Latif$^{11}$,
J.~Mammo$^{15}$,
Z.~S.~Meyers$^{12,7}$,
K.~Michaels$^{10}$,
K.~Mulrey$^{16}$,
A.~Nelles$^{12,7}$,
A.~Novikov$^{3}$,
A.~Nozdrina$^{3}$,
E.~Oberla$^{10}$,
B.~Oeyen$^{17}$,
Y.~Pan$^{18}$,
H.~Pandya$^{8}$,
I.~Plaisier$^{7,12}$,
N.~Punsuebsay$^{18}$,
L.~Pyras$^{12,7}$,
D.~Ryckbosch$^{17}$,
O.~Scholten$^{11,19}$,
D.~Seckel$^{18}$,
M.~F.~H.~Seikh$^{3}$,
D.~Smith\thanks{e-mail: danielsmith@uchicago.edu, authors@rno-g.org}$^{~,10}$,
D.~Southall$^{10}$,
J.~Torres$^{2}$,
S.~Toscano$^{1}$,
D.~Tosi$^{5}$,
D.~J.~Van Den Broeck$^{11,8}$,
N.~van Eijndhoven$^{11}$,
A.~G.~Vieregg$^{10}$,
C.~Welling$^{7,12}$,
D.~R.~Williams$^{20}$,
S.~Wissel$^{14,21}$,
R.~Young$^{3}$,
A.~Zink$^{7}$
}
\affiliation{
$^{1}$Universit\'e Libre de Bruxelles, Science Faculty CP230, B-1050 Brussels, Belgium\\
$^{2}$Dept. of Physics, Center for Cosmology and AstroParticle Physics, Ohio State University, Columbus, OH 43210, USA\\
$^{3}$University of Kansas, Dept.~of Physics and Astronomy, Lawrence, KS 66045, USA\\
$^{4}$National Nuclear Research University MEPhI, Kashirskoe Shosse 31, 115409, Moscow, Russia\\
$^{5}$Wisconsin IceCube Particle Astrophysics Center (WIPAC) and Dept.~of Physics, \\ University of Wisconsin-Madison, Madison, WI 53703,  USA\\
$^{6}$Uppsala University, Dept.~of Physics and Astronomy, Uppsala, SE-752 37, Sweden\\
$^{7}$Erlangen Center for Astroparticle Physics (ECAP), \\  Friedrich-Alexander-University Erlangen-Nuremberg, 91058 Erlangen, Germany\\
$^{8}$Vrije Universiteit Brussel, Astrophysical Institute, Pleinlaan 2, 1050 Brussels, Belgium\\
$^{9}$Dept.~of Physics and Astronomy, Michigan State University, East Lansing MI 48824, USA\\
$^{10}$Dept.~of Physics, Enrico Fermi Inst., Kavli Inst.~for Cosmological Physics, University of Chicago, Chicago, IL 60637, USA\\
$^{11}$Vrije Universiteit Brussel, Dienst ELEM, B-1050 Brussels, Belgium\\
$^{12}$DESY, Platanenallee 6, 15738 Zeuthen, Germany\\
$^{13}$Whittier College, Whittier, CA 90602, USA\\
$^{14}$Dept.~of Physics, Dept.~of Astronomy \& Astrophysics, Penn State University, University Park, PA 16801, USA\\
$^{15}$Dept.~of Physics and Astronomy, Univ.~of Nebraska-Lincoln, NE, 68588, USA\\
$^{16}$Department of Astrophysics/IMAPP, Radboud University, PO Box 9010, 6500 GL, The Netherlands\\
$^{17}$Ghent University, Dept. of Physics and Astronomy, B-9000 Gent, Belgium\\
$^{18}$Dept.~of Physics and Astronomy, University of Delaware, Newark, DE 19716, USA\\
$^{19}$Kapteijn Institute, University of Groningen, Groningen, The Netherlands\\
$^{20}$Dept.\ of Physics and Astronomy, University of Alabama, Tuscaloosa, AL 35487, USA\\
$^{21}$Physics Dept. California Polytechnic State University, San Luis Obispo CA 93407, USA\\
}

\begin{frontmatter}

\maketitle

\begin{abstract}
Over the last 25 years, radiowave detection of neutrino-generated signals, using cold polar ice as the neutrino target, has emerged as perhaps the most promising technique for detection of extragalactic ultra-high energy neutrinos (corresponding to neutrino energies in excess of 0.01 Joules, or $10^{17}$ electron volts). During the summer of 2021 and in tandem with the initial deployment of the Radio Neutrino Observatory in Greenland (RNO-G), we conducted radioglaciological measurements at Summit Station, Greenland to refine our understanding of the ice target. We report the result of one such measurement, the radio-frequency electric field attenuation length $L_\alpha$. We find an approximately linear dependence of $L_\alpha$ on frequency with the best fit of the average field attenuation for the upper 1500 m of ice: 
$\langle L_\alpha \rangle = \big( (1154 \pm 121) - (0.81 \pm 0.14) (\nu/\textnormal{MHz})\big)  \textnormal{ m}$ 
for frequencies $\nu \in [145 - 350] \textnormal{ MHz}$. 
\end{abstract}

\end{frontmatter}

\section{Introduction}

We report a measurement of the radio frequency electric field attenuation length of deep glacial ice at the US National Science Foundation's Summit Station in Greenland. This measurement is of interest to the ultra-high energy neutrino (UHEN) community due to the development of the Radio Neutrino Observatory in Greenland (RNO-G), a particle astrophysics experiment that uses the ice as a target material in the search for astrophysical and cosmogenic neutrinos \citep{RNO-G:2020rmc}. 

The IceCube experiment has placed a flux upper limit for astrophysical neutrinos of $E^2 \phi \lessapprox 2\times10^{-8} \textnormal{ GeV}/(\textnormal{cm}^2$ s sr) at $E_\nu=$1 EeV \citep{IceCube:2018fhm}. At such fluxes, a particle detector requires an active volume of O(10 km$^3$) or larger for a discovery-level detection within a detector's lifetime. A sparsely instrumented array of radio antennas, deployed in and on an extensive dielectric medium can satisfy this volume requirement. An UHEN interaction creates an extensive electromagnetic shower that produces impulsive radio emission via the Askaryan effect \citep{osti_4833087}. If the interaction occurs in an environment of low radio attenuation, a relatively small number of radio antennas can probe target volumes at the scale needed for UHEN observations. 

Glacial ice has been measured to have long radio attenuation lengths due to low temperature and relatively high purity  \citep{barwick_besson_gorham_saltzberg_2005,BESSON2008130,ALLISON2012457,Avva:2014ena,barrella_barwick_saltzberg_2011,hanson_2015}. This, combined with the volume of glacial ice available in Greenland and Antarctica, makes polar ice sheets attractive sites for the construction of a radio neutrino detector. 

RNO-G is one such experiment based on radio detection of UHEN in glacial ice \citep{RNO-G:2020rmc}, among others in Antarctica \citep{ALLISON2012457,ARA:2019wcf,ARIANNA:2019scz,RICE:2001ayk,ANITA:2008mzi,ANITA:2019wyx}. RNO-G is being constructed near NSF's Summit Station at the highest point of the Greenland Ice Sheet. The planned 
detector will ultimately be composed of 35 autonomous stations separated by 1.25 km in a grid pattern. Each station is instrumented with radio antennas, with good response over the range 100--600 MHz, deployed both just below the surface and at depths down to 100 m in boreholes.  Construction of the detector began during the summer of 2021 with the installation of the first three stations.

Previous measurements of radio attenuation lengths in Antarctica \citep{barwick_besson_gorham_saltzberg_2005,BESSON2008130,ALLISON2012457,barrella_barwick_saltzberg_2011,hanson_2015} and Greenland \citep{Avva:2014ena,macgregor_2015,paden1420297} have demonstrated that radio attenuation lengths vary at different ice locations, due primarily to differences in ice temperature and impurity levels. Since electric field attenuation length is a primary determinant of the expected number of observed UHENs at energies greater than 1 EeV, a precise, {\it in situ} measurement is required at Summit Station to assess RNO-G's science potential. 

Our work builds upon one previous {\it in situ} measurement of the bulk ice electric field attenuation length performed at Summit Station by \cite{Avva:2014ena}; that effort reported a depth-averaged attenuation length $\langle L_\alpha \rangle = 947^{+92}_{-85}$~m at 75 MHz. We herein quantify the attenuation length at higher frequencies to better match RNO-G's frequency range of 100-600 MHz. In addition to that prior analysis, 
there have been several previous measurements of the radio ice properties at Summit Station, including radar attenuation length measurements from air-borne radio sounding in the Greenland Ice Sheet Project \citep{macgregor_2015} and {\it in situ} radio sounding to investigate layering in the ice \citep{paden1420297}. We include a comparison of our reported attenuation with previous measurements at Summit Station and Antarctica in the \hyperref[sect:discussion]{\textbf{Discussion and Summary}} section. 

\section{Experimental Approach}

Our approach is similar to previous work in the astro-particle physics field \citep{Avva:2014ena,barrella_barwick_saltzberg_2011,BESSON2008130,hanson_2015,barwick_besson_gorham_saltzberg_2005}. 

We transmit an impulsive, broadband radio signal downwards into the ice via a wideband, directional antenna, and measure the return signal as a voltage versus time trace on a second, identical antenna. The transmitted signal propagates through the ice sheet, reflects off of the bedrock, and returns to the receiving antenna on the surface. After correcting for geometric path loss, bedrock coefficient of reflection, and electric field amplification from the focusing effect of the firn, the remaining power loss is attributed to absorption and scattering in the ice. Note that, experimentally, we do not distinguish between the two -- our quoted attenuation length implicitly includes both effects. 
(Possible dispersive effects at the bedrock are quantified in the \hyperref[sect:bedrockecho]{\textbf{Bedrock Echo}} section.) 
To reduce systematic uncertainties, we remove the system response of the electronics (initial impulse, antenna response, cables, amplifiers, filters) by normalizing against a second measurement run in air. We recreate the through-ice setup (same initial impulse, antenna polarization, cables, amplifiers, and filters but additional attenuators) for two antennas transmitting over a short distance in air and thereby largely cancel dependence on the system response from the in-ice data run. 

We take the reflection off the bedrock to be specular. Motivation for this decision is described in the \hyperref[sect:bedrockecho]{\textbf{Bedrock Echo}} section below. Given a specular reflection, the radar range equation reduces to the Friis equation \citep{friis}, and 
the direct through-air transmission formalism is applicable to our observed bedrock echoes. In that case,
following the notation used by \cite{Avva:2014ena}, the ratio of recorded voltages for each configuration (through-air vs. through-ice) as a function of frequency $\nu$ is,

\begin{equation}
    \frac{V_{\nu, ice}}{V_{\nu, air}} = \sqrt{F_{f} R} T_{ratio} \frac{d_{air}}{d_{ice}}  \exp\left(-\frac{d_{ice}}{\langle  L_\alpha \rangle}\right),\label{eq:1}
\end{equation}

\noindent where $d_{air}$ and $d_{ice}$ are the distances the ice and air signals travel between antennas, $R$ is the power reflection coefficient of the bedrock, $F_{f}$ is a focusing factor from the changing index of refraction in the firn \citep{Stockham2016RadioFI,phdthesis_stockham}, $T_{ratio}$ corrects for the change of transmission coefficient at the antenna feed between the antenna operating in air and in ice, and $\langle  L_\alpha \rangle$ is the depth-averaged electric field attenuation length over the entire depth of the ice. Solving for the attenuation gives,

\begin{equation}
    \langle  L_\alpha \rangle = d_{ice} / \ln \left( \sqrt{F_{f}R} T_{ratio} \frac{V_{\nu, air}}{V_{\nu, ice}}\frac{d_{air}}{d_{ice}} \right).
\end{equation}

This equation differs from that in \cite{Avva:2014ena} by the inclusion of the focusing factor, which arises from the amplification of field strength from propagation in the firn \citep{focusing_factor}. 
The addition of the focusing factor modifies the \cite{Avva:2014ena} bulk attenuation result from $947^{+92}_{-85}$~m to $913^{+85}_{-79}$~m.
We assume that the firn has an index of refraction varying linearly with the ice density profile $\rho(z)$ at Summit Station \citep{KOVACS1995245}. The density profile has been experimentally measured 
\citep{2013JGRF..118.1257A,hawley_morris_mcconnell_2008,alley_koci_1988,3a1027a412984298bf554c6237d562b4}
and fit to a double exponential \citep{Deaconu:2018bkf}, leading to a refractive index varying between $n\sim1.4$ at the surface and $n\sim1.78$ in deep ($<100$ m) ice. The changing index of refraction focuses power from the transmitter on the downwards path, such that the Fresnel zone radius at the bedrock is reduced relative to the constant refractive index case. After reflection at the bedrock, the signal is partially de-focused on the return path, however the electric field areal flux density at the surface receiver is still amplified compared to the $n$=constant case. The focusing factor is equivalent to a correction to the expected $1/R^2$ geometric spread factor, mathematically formalized by \cite{10.1111/j.1365-246X.1974.tb04105.x}. The focusing factor is present in \cite{macgregor_2015}, \cite{matsuoka_uratsuka_fujita_nishio_2004}, and \cite{Stockham2016RadioFI}, among others. We have used the finite-different time domain electrodynamics simulation software MEEP \citep{OSKOOI2010687} to confirm this effect. 

\section{Experimental Setup}

\begin{figure} 
\centering
\includegraphics[width=0.45\textwidth]{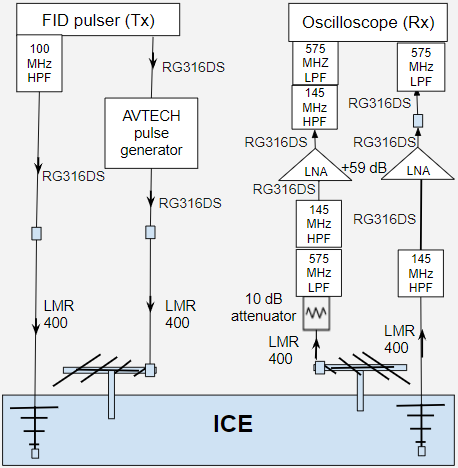}
\caption{Diagram of experimental setup for bedrock reflection. On the transmitting side, we use a self-triggering FID Technologies +5~kV high voltage pulse generator connected to a buried log-periodic dipole antenna (LPDA); an AVTECH fast pulse generator triggered by the FID pulser is connected to an in-air LPDA. On the receiving side, both the buried downward-pointing, and the in-air LPDAs are connected to a +59 dB low noise amplifier; those outputs are then recorded on a Tektronix digital oscilloscope, triggered by the in-air signal.}
\label{fig:GBS.png}
\end{figure}

A system diagram of the experiment is presented in Fig. \ref{fig:GBS.png}. This measurement was performed in August 2021 at Summit Station, Greenland, using a separation distance of 244 meters between the transmitter (coordinates 72.5801$^\circ$N, 38.4569$^\circ$W) and receiver (coordinates 72.5786$^\circ$N, 38.4527$^\circ$W) sites. 
The large separation distance assured that direct propagation from the transmitting antenna to the receiving antenna did not saturate the receiving amplifier. 
All antennas used were commercially available Create CLP-5130-2N \citep{CREATE_MANUAL} log-periodic dipole antennas (LPDA) with $\sim 8$ dBi in-air forward gain over the band 105--1300 MHz. 

Due to the large distance between stations, two parallel electronics signal chains are used, for triggering and the bedrock echo measurement, respectively. 
The bedrock echo electronics signal chain starts with the self-triggered high voltage FID Technology\footnote{http://www.fidtechnology.com} model FPG6-1PNK pulse generator, which delivers a +5~kV signal to a $50\Omega$ coaxial feed. After the FID output, we apply a 100 MHz high pass filter using a Minicircuits\footnote{https://www.minicircuits.com/products/RF-Filters.html} NHP-100 filter. 
Following the filter, the signal is conveyed over 12 m of LMR-400 $50 \Omega$ coaxial cable to an LPDA buried in the ice and pointed vertically downwards towards the bedrock; the bedrock-reflected return signal is then measured by a similarly-buried, downwards-pointing, receiver LPDA.  The receiving antenna, located 244~m away along the surface of the ice, is aligned with the antenna tines parallel and collinear to the ones of the transmitting antenna, so that each antenna is in the gain null of the other to minimize contamination from horizontal ray paths. After measurement in the receiving antenna, the signal travels over a 10 m LMR-400 cable, bandpass filtered from 145-575 MHz using Minicircuits VHF-145+ and VLF-575+ filters, and then amplified by a custom RNO-G design low-noise amplifier with +59 dB of gain over the band 80--750~MHz. After the amplifier, the signal is bandpass filtered again using Minicircuits NHP-200 and VLF-575+ filters, and then recorded on a 2 GHz-bandwidth Tektronix MSO5204B oscilloscope. 

The oscilloscope is triggered by the second electronic signal chain, ensuring a stable trigger over the distance between transmitter and receiver. 
The second chain begins with an AVTECH AVIR-1-C pulse generator triggered by the FID pulse generator \texttt{TRG OUT}, producing an impulsive, O(1 ns) pulse. The pulse generator is connected over a 12 m LMR-400 cable to an elevated, in-air LPDA pointed at a similarly elevated receiver LPDA located 244~m away and viewing the transmitter on boresight. The received in-air signal is then attenuated by 20 dB to prevent saturation, bandpass filtered using Minicircuits VHF-145+ and VLF-575+ filters, amplified by +59 dB using the RNO-G low-noise amplifier, bandpass filtered again using Minicircuits VHF-145+ and VLF-575+ filters and finally captured by the oscilloscope. This in-air signal was used to trigger the oscilloscope, and therefore provides the reference $t_0$ for our measurements.

The oscilloscope was set to collect data over a 50 $\mu s$ window; 10,000 individual triggers are averaged to suppress incoherent noise contributions and that average is written to scope memory. Twenty 10,000-event runs were collected, and then again averaged in post-processing, bringing the total number of triggers to 200,000. 

To perform the air $\to$ air normalization run, we swapped the cables for the in-ice antennas with those from the in-air antennas. On the receiving side, two modifications were made: we added a 46 dB attenuator to prevent amplifier saturation and, for this configuration, we self-triggered on the arriving signal. 

\section{Experimental Results}

\subsection{Bedrock Power Reflection Coefficient}

The power reflection coefficient at the ice-rock interface is not well-known and constitutes the largest uncertainty in our measurement of attenuation length. Taking an approach similar to \cite{Avva:2014ena}, we take the power reflection coefficient to have a mean value of 0.215, a typical value for ice-bedrock interfaces as derived from radio sounding experiments \citep{chris_reflect,barwick_besson_gorham_saltzberg_2005}. For uncertainty analysis, we assume the reflection coefficient can be drawn from a probability density function, uniformly distributed in the log of the reflection coefficient over the range from 0.01 to 1.0, which represent plausible extrema for the interface, from a frozen bedrock with high water content to an underlying layer of water \citep{chris_reflect}. 

The observed return echo, in principle, could include both coherent and also incoherent contributions. Whereas the former sum linearly with the number of average triggers, the latter will scale as the square root of the number of events averaged $N_{avg}$ \citep{paden_thesis}. We have explicitly verified that our final results are insensitive to $N_{avg}$, consistent with the assumption that the observed specular return echo, after subtracting the contribution from noise, is dominated by coherent scattering. 

\subsection{Bedrock Echo}
\label{sect:bedrockecho}

\begin{figure*}[ht]

\begin{subfigure}{.65\textwidth}
  \centering
  \includegraphics[width=1.0\linewidth]{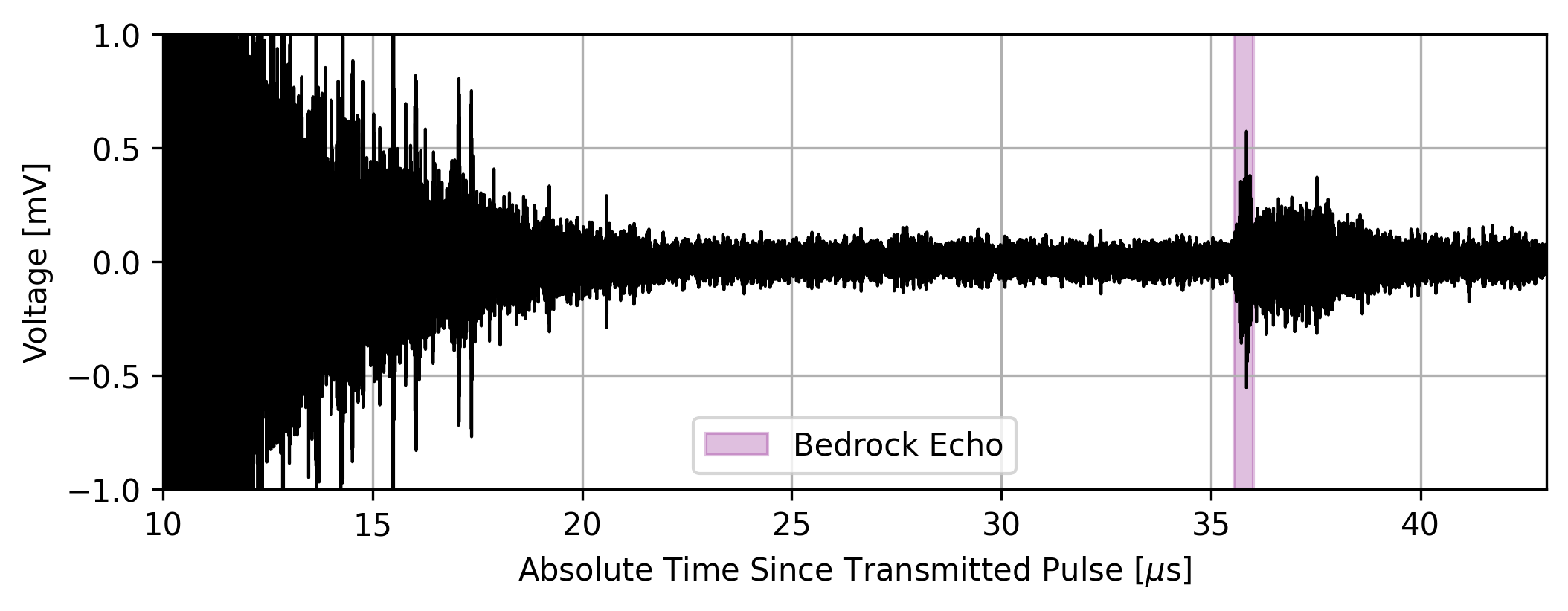}
\end{subfigure}%
\begin{subfigure}{.35\textwidth}
  \centering
  \includegraphics[width=1.0\linewidth]{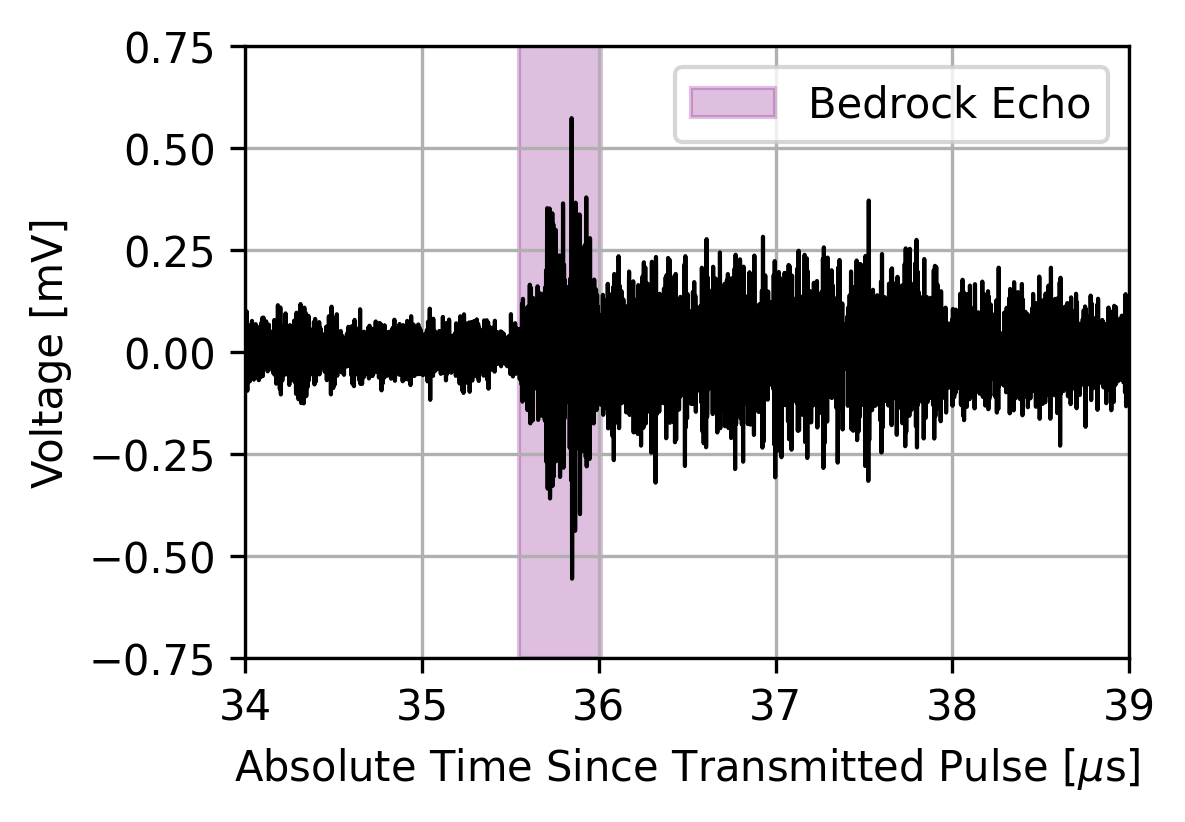}
\end{subfigure}

\begin{subfigure}{.65\textwidth}
  \centering
  \includegraphics[width=1.0\linewidth]{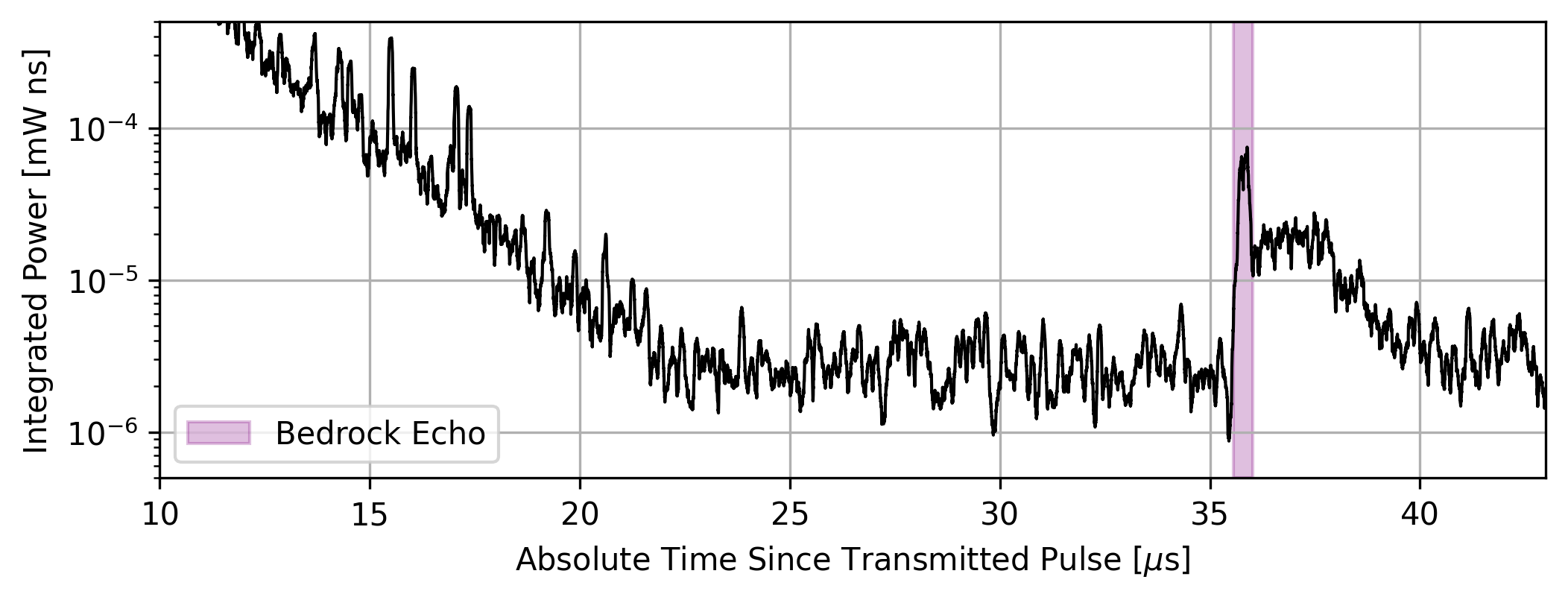}
\end{subfigure}%
\begin{subfigure}{.35\textwidth}
  \centering
  \includegraphics[width=1.0\linewidth]{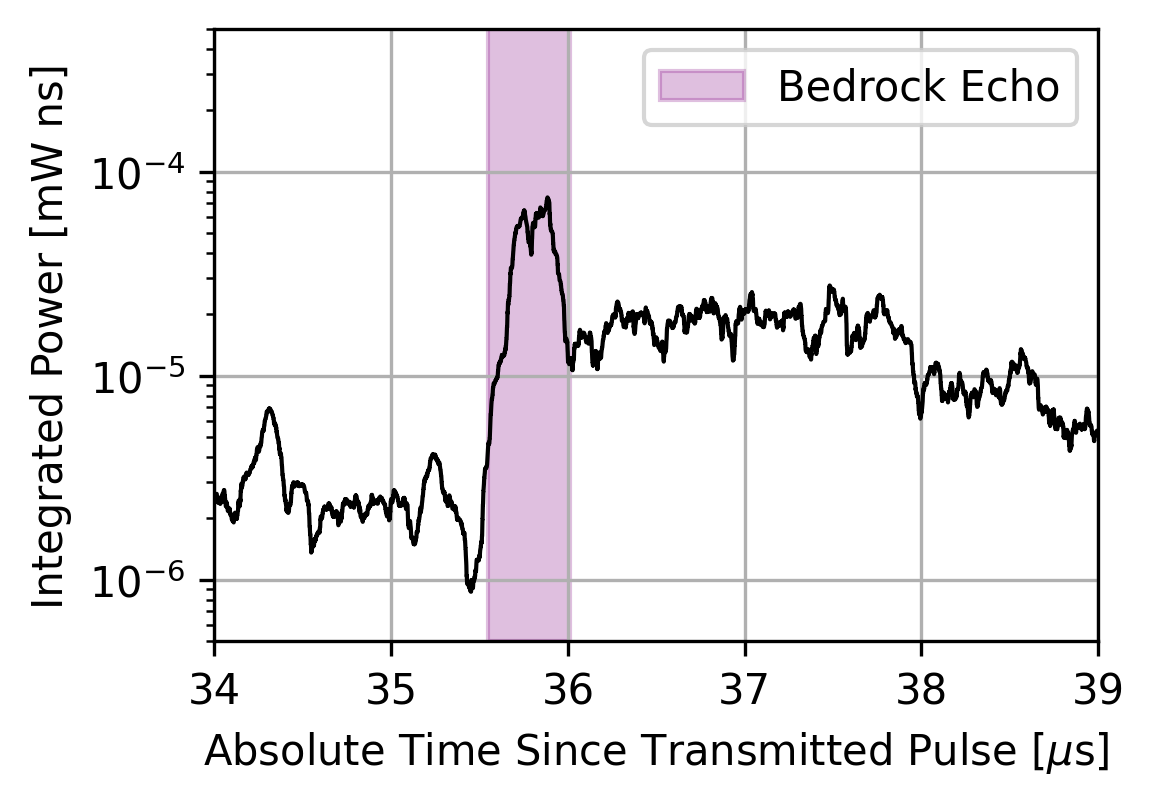}
\end{subfigure}

\caption{Top: Recorded voltage as a function of time for the receiving in-ice antenna. 
Bottom: Recorded power, integrated in a sliding window of 100 ns to account for the group delay of the LPDA antennas.  
The specular component of the bedrock echo `signal' is highlighted in magenta. Sub-surface internal layer reflections are visible at times earlier than 22 $\mu s$, after which noise dominates up to the point at which the bedrock echo is evident. }
\label{fig:gb_time}
\end{figure*}

The reflection from the bedrock is visible above thermal noise
in the time domain voltage trace of the receiving antenna, at a signal onset time of 35.55 $\mu s$ after the oscilloscope trigger (Fig. \ref{fig:gb_time}). The bedrock echo is observed to include two components: a predominantly-specular, sharp, faster impulse (of duration $\sim500 ns$), and a long ($>2 \mu s$) extended signal which we associate with more diffuse, multi-path reflections off irregular features, both on the surface of, and within, the underlying bed reflector. For the purposes of the bulk radio attenuation measurement, and since an extended tail is not present for the in-air normalization run, we restrict consideration to the fast, specular component. 

The uncertainty in the time window of the specular reflection is of $O(10 ns)$, dominated by noise fluctuations at the edges of the window. This uncertainty is neglected since it is sub-dominant relative to the other systematic uncertainties in our final measurement. The final window start and end times are therefore defined to be 35.55 and 36.05 $\mu s$, respectively. 

\begin{figure}
\centering
\includegraphics[width=0.48\textwidth]{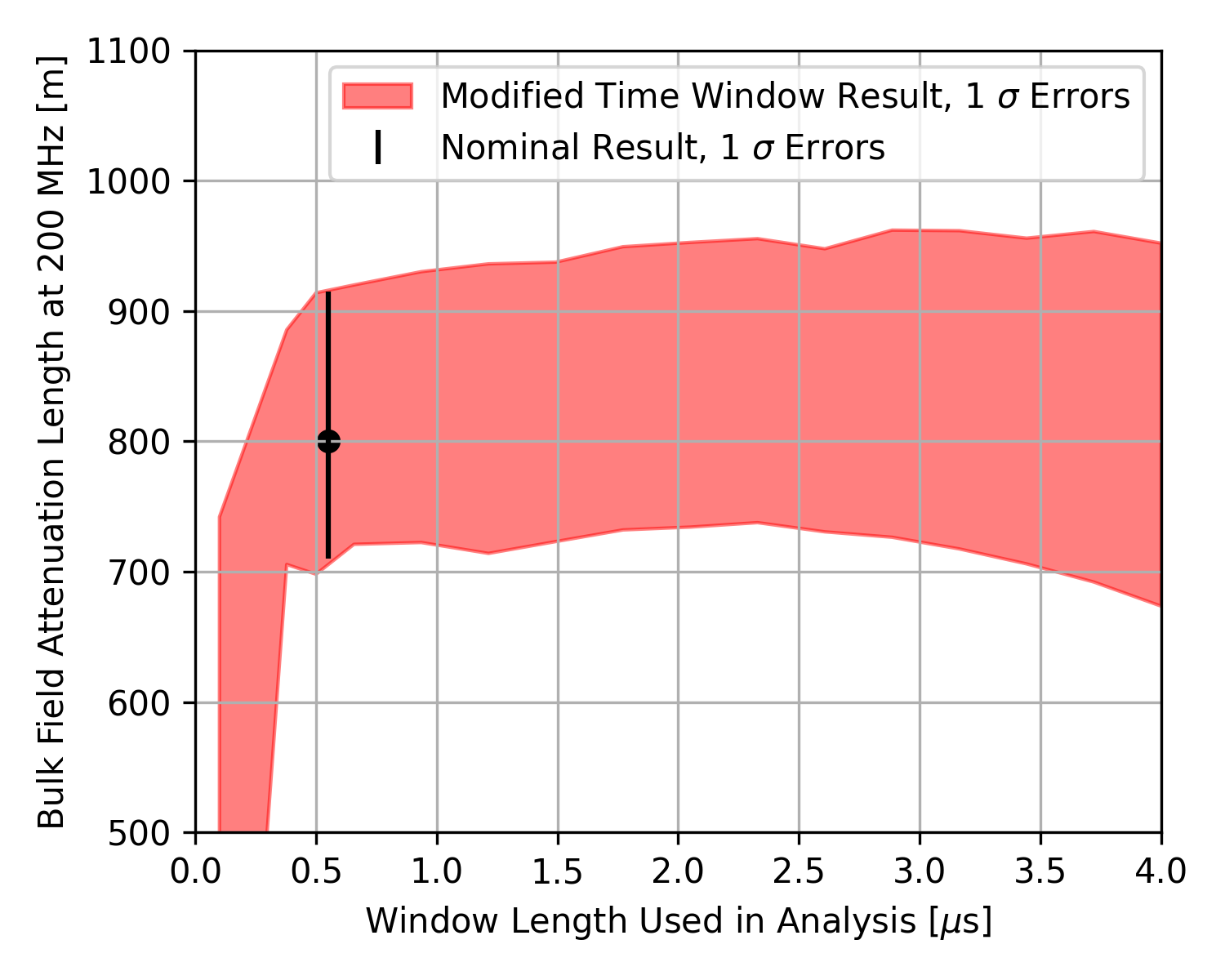}
\caption{Measurement of the depth-averaged electric field attenuation at 200 MHz as a function of the window length used to select the bedrock echo. We note that including the entire diffuse component of the reflection into the final attenuation calculation increases the final result by no more than 10\%. } 
\label{fig:increased_window}
\end{figure}

To determine the impact of neglecting the diffuse component of the bedrock echo on our final measured value of bulk attenuation, we investigate the dependence of our numerical result on the window length used in our analysis. We expect the measured attenuation to increase with increased window length due to the extended integration of power returning from the bedrock; at frequencies below 250~MHz, we obtain a $\sim$10\% larger attenuation length, but with increased uncertainty, as seen in Fig. \ref{fig:increased_window}. At frequencies above 250~MHz, there is a negligible increase in attenuation length with increased window length. We note that the additional attenuation length is within the systematic uncertainties of our stated result due to the large bedrock reflection coefficient uncertainty. 

The relationship between bulk ice attenuation and received power is different for the specular vs. diffuse components. We define attenuation from the Friis transmission equation (implicit in Eq. \ref{eq:1}) \citep{friis}. The Friis transmission equation is applicable for a specular reflection as it assumes direct line-of-sight propagation without interference within the first Fresnel zone, leading to a geometric path loss $\propto d_{ice}^2$. The radar range equation is more applicable to the diffuse component since it includes power contributions from a rough surface via the definition of a radar cross section and a geometric path loss $\propto (d_{ice}/2)^4$ \citep{Balanis-2012-antenna}. We find that use of the radar range equation instead of the Friis equation over the combined specular and diffuse components also increases the measured attenuation length at lower frequencies by a maximum of 10\%, albeit introducing more model dependence from the unknown value of bedrock radar cross section. 

Based on the relatively small increase in obtained attenuation length from including the diffuse component, we quote our final result based on the Friis equation; this choice is consistent with previous similar measurements \citep{barwick_besson_gorham_saltzberg_2005,BESSON2008130,barrella_barwick_saltzberg_2011,hanson_2015,Avva:2014ena,paden1420297}. 

\subsection{Bedrock Depth}
\label{sect:bedrockdepth}

The bedrock depth can be derived from the absolute time of flight of the transmitted pulse and a model for the index of refraction of the ice as a function of depth. 
To reduce systematic biases (from location extrapolation and bedrock radio properties) and also as a cross check of our absolute timing calibration, we measure the bedrock depth from these data, rather than relying on previous measurements .
The relationship between time of flight ($\Delta t$) and bedrock depth (half of the total distance propagated by the transmitted signal, $d_{ice}$) can be found by solving for $d_{ice}$ in the integral:

\begin{equation}
\label{eq:tof}
    \Delta t = \frac{2}{c} \int_0^{d_{ice} / 2} n(z) dz,
\end{equation}

\noindent where $n(z)$ is the model for the index of refraction as a function of depth [m]. Index of refraction is related to dielectric constant ($\epsilon'$) via $n(z) = \sqrt{\epsilon'(z)}$. The value of $\epsilon'(z)$ is derived from its relationship with measured ice density ($\rho$, [kg / m$^3$]) \citep{KOVACS1995245,Barwick:2018rsp}: 

\begin{equation}
\label{eq:permittivity}
    \epsilon'(z) = \left(1 + 0.854 \rho(z)\right)^2.
\end{equation}

The parameterization of the dependence of ice density on depth follows \cite{Deaconu:2018bkf}, who performed a double exponential fit with a critical density at a depth of 14.9 m \citep{herron_langway_1980}:

\begin{equation}
\rho(z) = 
\begin{cases}
    0.917 - 0.594 e^{-z / 30.8} & z \leq 14.9~\textnormal{m} \\
    0.917 - 0.367 e^{-(z - 14.9) / 40.5} & z > 14.9~\textnormal{m} \\
\end{cases}
\end{equation}

The uncertainty on the depth determination arises primarily from the uncertainty in the asymptotic index of refraction of deep glacial ice, which we take to be $n = 1.78 \pm 0.03$ \citep{Bogorodsky_1985} for ice below the firn (deeper than 100 m at Summit Station). Using this refractive index profile, we calculate the bedrock to be at a depth of $3004^{+50}_{-52}$~m, corresponding to $d_{ice} = 6008^{+100}_{-104}$~m for the through-ice bedrock echo total travel distance. 

We note that, while our transmitting and receiving antennas were separated by 244~m, the through-ice signal approximately propagates vertically and Eq. \ref{eq:tof} holds true. Over the measured $6$~km propagation distance, horizontal propagation results in 4~m extra path length. 

This bedrock depth is consistent with previous measurements of the bedrock depth from \cite{Greenland_Ice_Sheet_Project_1994}, at $3053.5$~m in 1993, and from \cite{Avva:2014ena}, at $3014^{+48}_{-50}$~m in 2015. 

\subsection{Antenna Coupling}

\begin{figure}
\centering
\includegraphics[width=0.48\textwidth]{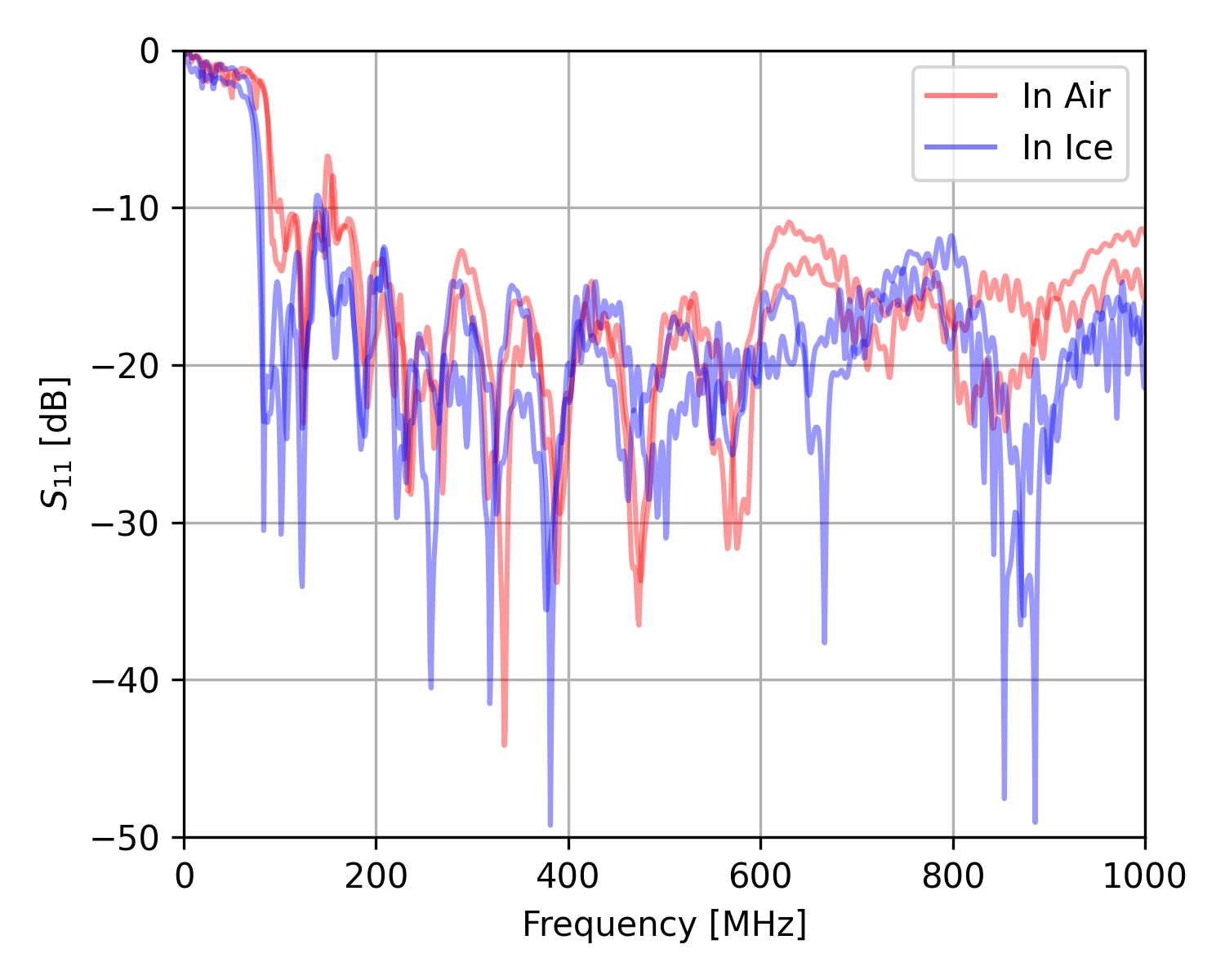}
\caption{Measured $S_{11}$ of each antenna used in the experiment. The difference in the low-frequency cutoff of the antenna when it is embedded in the ice compared to in air is due to the different indices of refraction of the two environments.}
\label{fig:s11}
\end{figure}

The antenna transmission coefficient is defined as the quantity of power transmitted by the antenna from an incident radio frequency signal on a 50 $\Omega$ transmission line at the antenna feed point ($S_{21}$ in the scattering matrix). The transmission coefficient depends upon the dielectric properties of the antenna's embedded environment \citep{nuradioreco2019,Barwick:2014boa}. 
To increase the power transmitted into the ice, we buried our antennas so that all active conductors were at least $\sim20$ cm below the surface, thereby embedding them in an environment of $n \approx 1.4$. 
The antennas used for the normalization are in air, for which $n \sim 1.0$. To correct for the change in match, we calculate a $T_{ratio}$ from the measured reflection coefficient ($S_{11}$ as shown in Fig. \ref{fig:s11}) of the four antennas, two in air and two in ice, taken in the field. Assuming that all power not reflected at the feed is transmitted, the ratio becomes, in terms of the reflection coefficient $S_{11}$ in dB,

\begin{equation}
    T_{ratio} = \frac{1 - 10^{S_{11, ice} / 10}}{1 - 10^{S_{11, air} / 10}}.
\end{equation}

The antennas were found to transmit nearly all incident power in the frequency range of interest (150--550 MHz) in both the in-air and in-ice cases, resulting in a small $T_{ratio}$ correction. Averaged over frequency in the range of interest, $T_{ratio} = 1.00 \pm 0.05$, with the uncertainty assessed empirically from the variance of the measured match over the band. 
Our result is consistent with $T_{ratio}$ measured by other groups using the same or similar antennas \citep{barrella_barwick_saltzberg_2011,Avva:2014ena}.

{\it In situ} measurements, as well as simulations \citep{Barwick:2014boa,nuradioreco2019}, have shown that the frequency-dependent antenna gain $G_0$, measured in air, also changes when the antenna is embedded in a dielectric medium. This change can be modeled as a down-shift in frequency, by the index of refraction of the medium ($G'(\nu) = G_0(\nu / n)$). For the LPDA used in this work, the gain over the frequency band of interest is uniform to $<0.5$~dBi \citep{CREATE_MANUAL}, rendering the shift between in-air and in-ice measurements a subdominant systematic bias. The down-shift in frequency will cause a corresponding shift in low-frequency cutoff both in the gain and in the $S_{11}$ (as shown in Fig. \ref{fig:s11}), but the cutoff in both environments is below the high-pass filter of our analysis. 

We note that there will be different contributions from the ice surface in both the in-air and in-ice antenna responses. We neglect these effects because they are likely to be small so long as the directional antennas (front-to-back [F/B] ratio of the LPDA $\sim$-15 dB) are pointed away from the surface \cite{Barwick:2014boa}.

\subsection{Firn Focusing}

There is a geometric amplification of the bedrock echo electric field at the surface of the ice due to the changing index of refraction of the firn. To calculate the power focusing factor, for a negligibly thin firn layer, straightforward application of Snell's law prescribes that the electric field flux density measured at the surface, after bedrock reflection, follows:\citep{Stockham2016RadioFI}:

\begin{equation}
    F_f = 
    \left(\frac{n(z = 0)}{n(z = d_{ice} / 2)}\right)^2.
\end{equation}

We have verified that this equation agrees with a ray tracing simulation and 3D FDTD simulation. The uncertainty on the focusing factor arises from the uncertainty in the index of refraction model. Using $n = 1.78 \pm 0.03$ \citep{Bogorodsky_1985} (as described previously in the \hyperref[sect:bedrockdepth]{\textbf{Bedrock Depth}} section), and  $n = 1.4 \pm 0.1$ for the surface ice \citep{Bogorodsky_1985}, we obtain a final focusing factor of $F_f = 1.61 \pm 0.24$. 

\subsection{In-Air Normalization Amplitude}

\begin{figure}
\centering
\includegraphics[width=0.48\textwidth]{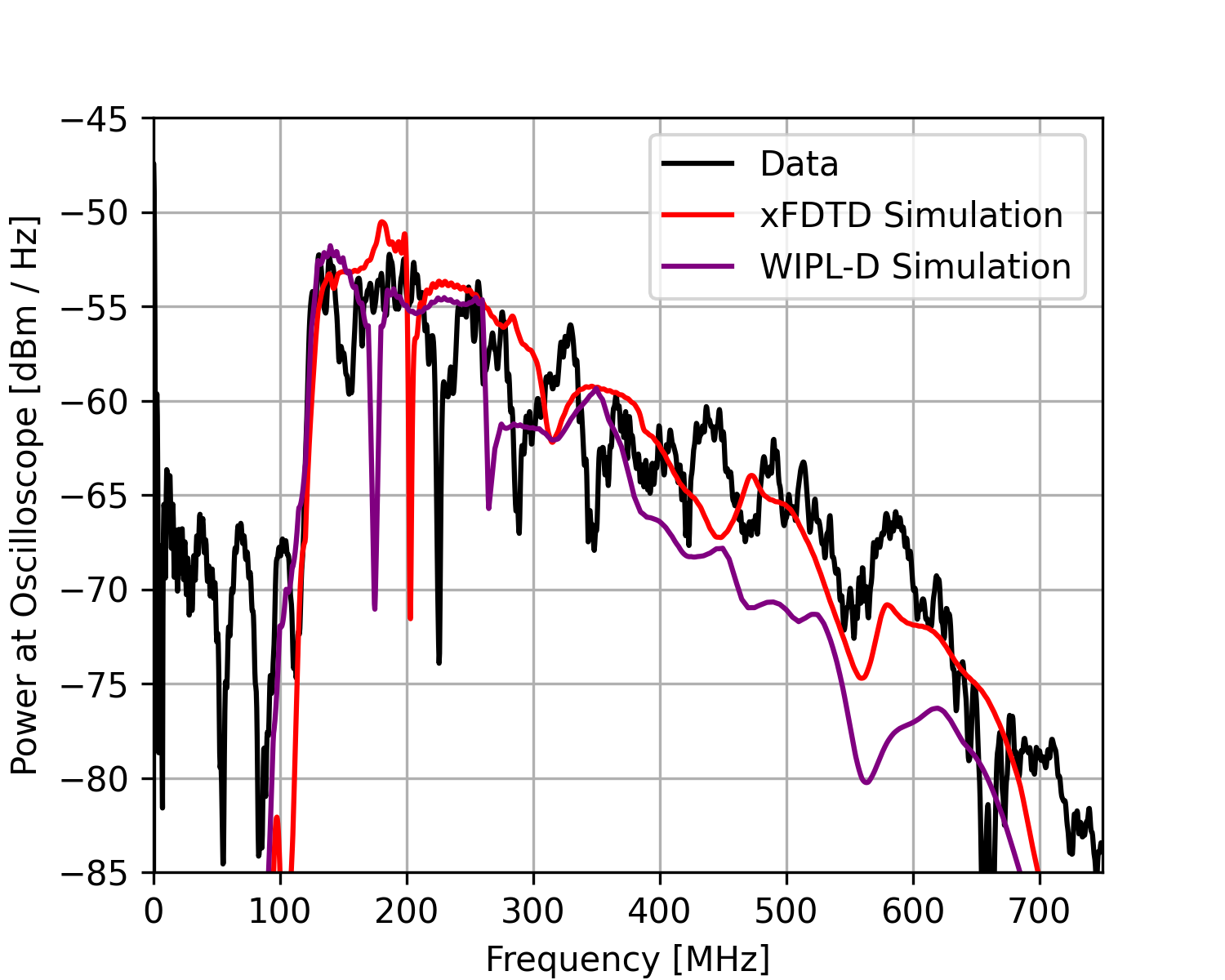}
\caption{Comparison of data from the in-air normalization run compared against an absolute amplitude expectation as derived from two separate LPDA simulations. No systematic bias is evident within the $\pm$10\% voltage uncertainty in the antenna model \citep{Barwick:2016mxm,Barwick:2014boa}, over the frequency band of this analysis. The sharp dip between 180--220~MHz seen in data and simulations is most likely due to fine details in tine length and separation, which may be difficult to  accurately simulate \citep{Barwick:2016mxm}.
}
\label{fig:norm}
\end{figure}

The amplitude of the signal from the in-air normalization run can be systematically biased from reflections off of the ice surface, increasing or decreasing the recorded power observed from the direct line-of-sight signal. Given the antenna heights above the ice (1.5~m) and distance between antennas (244 m), the first Fresnel zone is comprised by a nearly uniform, planar surface ice reflector, at all frequencies of interest. This leads to potential interference from reflections, depending on geometry: direct rays will interfere destructively/constructively with rays at the center/periphery of the Fresnel zone.  To quantify any possible systematic bias, we compare the data against the absolute amplitude expectation of the signal from simulation. The absolute amplitude is derived from a measurement of the FID pulse shape, amplifier response, filter response, free space path loss, and two independent simulations of the LPDA antenna response. Our simulations use either the Method of Moments software WIPL-D\footnote{\url{https://wipl-d.com}} or Finite Difference Time Domain software xFDTD\footnote{\url{https://www.remcom.com/xfdtd-3d-em-simulation-software}}, and have been found to agree with anechoic chamber measurements to 10\% uncertainty \citep{Barwick:2016mxm,Barwick:2014boa}. The comparison of the simulated result with our data, seen in Fig. \ref{fig:norm}, demonstrates that any possible systematic bias is not greater than 10\% in voltage, consistent with previous results \citep{Barwick:2016mxm,Barwick:2014boa}.

\subsection{Result and Error Analysis}

\begin{figure}
\centering
\includegraphics[width=0.48\textwidth]{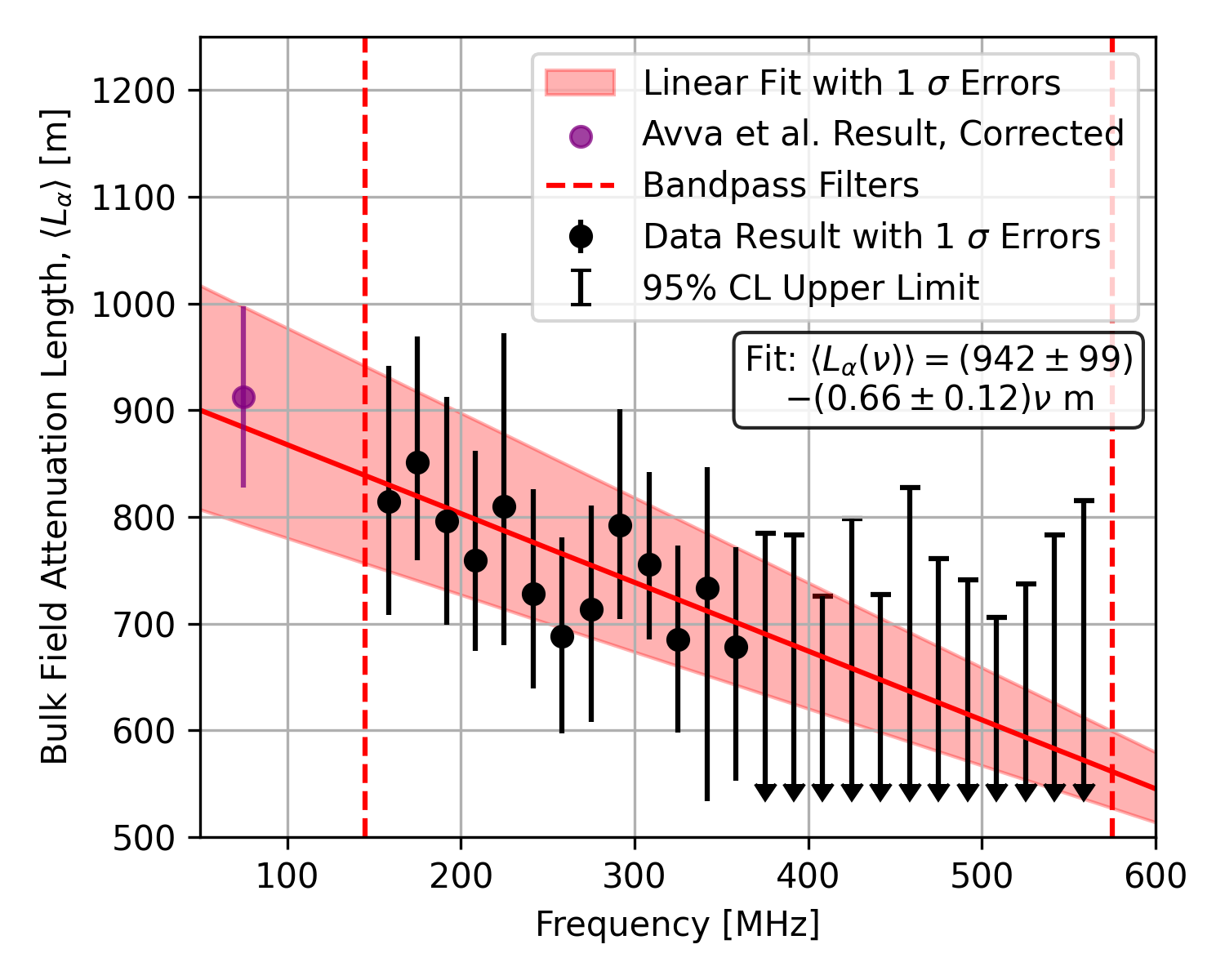}
\caption{Measurement of the depth-averaged electric field attenuation as a function of frequency at Summit Station, within the system bandpass (shown as red dashed lines).  Definitions of the error bars (and displayed upper limits) are provided in the text. The corrected result from \cite{Avva:2014ena} is shown for comparison. The frequency $\nu$ in the fit has units of MHz.} 
\label{fig:result}
\end{figure}

The measured depth-averaged field attenuation length is presented in Fig. \ref{fig:result}. The previous {\it in situ} measurement reported by \cite{Avva:2014ena} is included for comparison, adjusted to remove the systematic bias from firn focusing not previously included in that analysis, resulting in a correction of their published result (from $947^{+92}_{-85}$m to $913^{+85}_{-79}$m). We report the measurement only within the bandpass limits of our system, over which we have the highest sensitivity and lowest systematic biases from antenna modeling and filter response. Beyond the system bandpass limits, we have checked that our procedure yields an attenuation length numerically consistent with zero as expected for a noise-dominated regime.

Contributions of different sources of uncertainty are calculated using a Monte Carlo method. We numerically calculate the estimated probability density distribution (PDF) of the bulk electric field attenuation within each frequency bin by repeatedly drawing random values of each component of the final measurement from their respective PDFs. 
For systematic uncertainties, we assume that each measured quantity used in the calculation of bulk field attenuation is uncorrelated and has a PDF either of a normal distribution (as is the case for $F_f$, $T_{ratio}$ and $d_{ice}$) or the distribution already described in the text (as is the case for $R$). 
The main component of statistical uncertainty is due to fluctuations in the power contributions from thermal noise in the recorded oscilloscope trace. In the 150-300~MHz band, the uncertainty from noise statistical fluctuation is sub-dominant to systematic uncertainties, contributing less than $10\%$ to the quoted uncertainty of each frequency bin.
The final measurement is reported as a central value with one standard deviation (statistical plus systematic) error bars for those frequency bins that yield statistically significant results. For all other frequency bins, we report a 95\% confidence level upper limit.

 It is important to note that the majority of uncertainties are correlated in each frequency bin of the final measurement, with the primary contributions to the uncertainty arising from finite noise statistics and small systematic biases from the difference in LPDA response in ice vs. air. For the linear fit presented below, the visibly high goodness of fit is due to this correlation of uncertainties between frequency bins. 

The reported bulk attenuation length includes losses from layer scattering. While expected to be a subdominant effect for vertical propagation due to the low reflection coefficient of the observed layers \citep{paden1420297}, quantifying effects due to layer scattering for the more horizontal neutrino geometries must account for the larger Fresnel reflection coefficients at more glancing layer incidence angles \citep{christoph_energy}.

Birefringence of the ice can result in rotation of the signal that is dependent on polarization, leading to apparent loss of power at the co-polarized receiver.
Measurements of the crystal orientation at Summit Station indicate uniaxial fabric at all depths \citep{thorsteinsson1996textures}, unlike South Pole \citep{Matsuoka2003CrystalOF,barwick_besson_gorham_saltzberg_2005}, indicating that birefringence will matter less for the Greenland site, though this remains to be quantified.

\subsection{Discussion and Summary}
\label{sect:discussion}

We derive the electric field attenuation length as a function of depth using a model of the field attenuation's dependence on temperature and chemical impurities, as described in detail  in \cite{macgregor_2007, macgregor_2015, wolff}. For a medium with non-zero intrinsic conductivity, the attenuation length can be expressed as:

\begin{equation}
    L_{\alpha}(f, z) = A(f) \frac{\epsilon_0 \sqrt{\epsilon'_r(z)} c}{\sigma_{\infty}(z)},
\end{equation}

\noindent where $c$ is the speed of light, $\epsilon_0$ is the permittivity of free space, $\epsilon'_r(z)$ is the real component of the relative permittivity at a given depth $z$ (defined in Eq. \ref{eq:permittivity}), $\sigma_{\infty}(z)$ is the infinite frequency limit of the electrical conductivity at a given depth, and the parameter $A(f)$ is extracted experimentally, by requiring that the integrated, depth-dependent attenuation match our measured value of full-path attenuation, at each frequency $f$. \cite{fujita2000summary} suggest that the infinite frequency limit conductivity is valid at radio frequencies, since the molar conductivity does not change from low frequency (the frequency at which reference conductivity was measured, 0.3-3~MHz, \citep{wolff}) to our frequency band (VHF/UHF, 150-550~MHz). The infinity frequency conductivity is related to chemical impurities and temperature via,

\begin{equation}
\begin{split}
    \sigma_\infty(z) & = 
    \sigma_{\text{pure}} \exp \left[ \frac{E_{\text{pure}}}{k} \left(\frac{1}{T_r} - \frac{1}{T(z)}\right)\right] \\
    & + \mu_{\text{H}^+} [\text{H}^+](z) \exp \left[ \frac{E_{\text{H}^+}}{k} \left(\frac{1}{T_r} - \frac{1}{T(z)}\right)\right] \\
    & +\mu_{\text{Cl}^-} [\text{Cl}^-](z) \exp \left[ \frac{E_{\text{Cl}^-}}{k} \left(\frac{1}{T_r} - \frac{1}{T(z)}\right)\right] \\
    & + \mu_{\text{NH}_4^+} [\text{NH}_4^+](z) \exp \left[ \frac{E_{\text{NH}_4^+}}{k} \left(\frac{1}{T_r} - \frac{1}{T(z)}\right)\right],
\end{split}
\end{equation}

\noindent where $k$ is Boltzmann's constant, $T(z)$ is in-ice temperature at a given depth as measured at the GRIP borehole \cite{Greenland_Ice_Sheet_Project_1994}\footnote{\label{ftp_temp}\url{ftp://ftp.ncdc.noaa.gov/pub/data/paleo/icecore/greenland/summit/grip/physical/griptemp.txt}}, $T_r$ is a reference temperature, $\sigma_{\text{pure}}$ is the conductivity of pure ice, $\mu_{\text{H}^+}$, $\mu_{\text{Cl}^-}$, and $\mu_{\text{NH}_4^+}$ are molar conductivities, $E_{\text{pure}}$, $E_{\text{H}^+}$, $E_{\text{Cl}^-}$, and $E_{\text{NH}_4^+}$ are activation energies, and $[\text{H}^+](z)$\footnote{\label{ftp_dep}\url{ftp://ftp.ncdc.noaa.gov/pub/data/paleo/icecore/greenland/summit/grip/ecm/gripdep.txt}}, $[\text{Cl}^-](z)$\footnote{\label{ftp_chem}\url{ftp://ftp.ncdc.noaa.gov/pub/data/paleo/icecore/greenland/summit/grip/chem/gripion.txt}} and $[\text{NH}_4^+](z)^\text{\ref{ftp_chem}}$ are depth dependent molar concentrations as measured at the GRIP borehole \citep{Greenland_Ice_Sheet_Project_1994, LEGRAND1993251, h_plus_description}. The values of molar conductivities, conductivity of free ice, and activation energies as measured by \cite{macgregor_2007, macgregor_2015} and used in this analysis are given in Table \ref{tbl:parameters}. Note that, in this formulation, the temperature dependence of the attenuation length is explicitly absorbed into the conductivity dependence on temperature - the zero conductivity limit would correspond to no absorption and, correspondingly, no attenuation, for any temperature.

\begin{table}
\begin{center}
\begin{tabular}{c c c c} 
 Symbol & Description & Unit & Value \\ [0.5ex] 
 \hline\hline
$T_r$ & Reference temperature & $^\circ$C & -21 \\ 
 \hline
$\sigma_\text{pure}$ & Conductivity of pure ice & $\mu$S/m & $9.2\pm0.2$ \\ 
 \hline
$\mu_{\text{H}^+}$ & Molar conductivity of H$^+$ & S/m/M & $3.2\pm0.2$ \\  
 \hline
 $\mu_{\text{Cl}^-}$ & Molar conductivity of Cl$^-$ & S/m/M & $0.43\pm0.07$ \\ 
 \hline
 $\mu_{\text{NH}_4^+}$ & Molar conductivity of NH$_4^+$ & S/m/M & $0.8$ \\ 
 \hline
 $E_{\text{pure}}$ & Activation energy of pure ice & eV & $0.51\pm0.01$ \\ 
  \hline
 $E_{\text{H}^+}$ & Activation energy of H$^+$ & eV & $0.20\pm0.04$ \\ 
  \hline
 $E_{\text{Cl}^-}$ & Activation energy of Cl$^-$ & eV & $0.19\pm0.02$ \\ 
  \hline
 $E_{\text{NH}_4^+}$ & Activation energy of NH$_4^+$ & eV & $0.23$ \\ 
 \hline
 T(z) & Ice temperature at depth $z$ & $^\circ$C &  See \ref{ftp_temp}  \\ 
\hline 
$[\text{H}^+](z)$ & Molar concentration of H$^+$ & $\mu$M & See \ref{ftp_dep} \\ 
\hline 
$[\text{Cl}^-](z)$ & Molar concentration of Cl$^-$ & $\mu$M & See \ref{ftp_chem}  \\ 
\hline 
$[\text{NH}_4^+](z)$ & Molar concentration of NH$_4^+$ & $\mu$M & See \ref{ftp_chem}  \\ 
\end{tabular}
\caption{Values of parameters used in the conductivity model of ice at Summit Station. Compiled from \cite{macgregor_2007, macgregor_2015} and from the GRIP borehole \citep{Greenland_Ice_Sheet_Project_1994}. Molar concentrations and ice temperature are tables of data measured at the GRIP borehole, and are available at the corresponding links in the footnotes.} 
\label{tbl:parameters}
\end{center}
\end{table}

\begin{figure}
\centering
\includegraphics[width=0.45\textwidth]{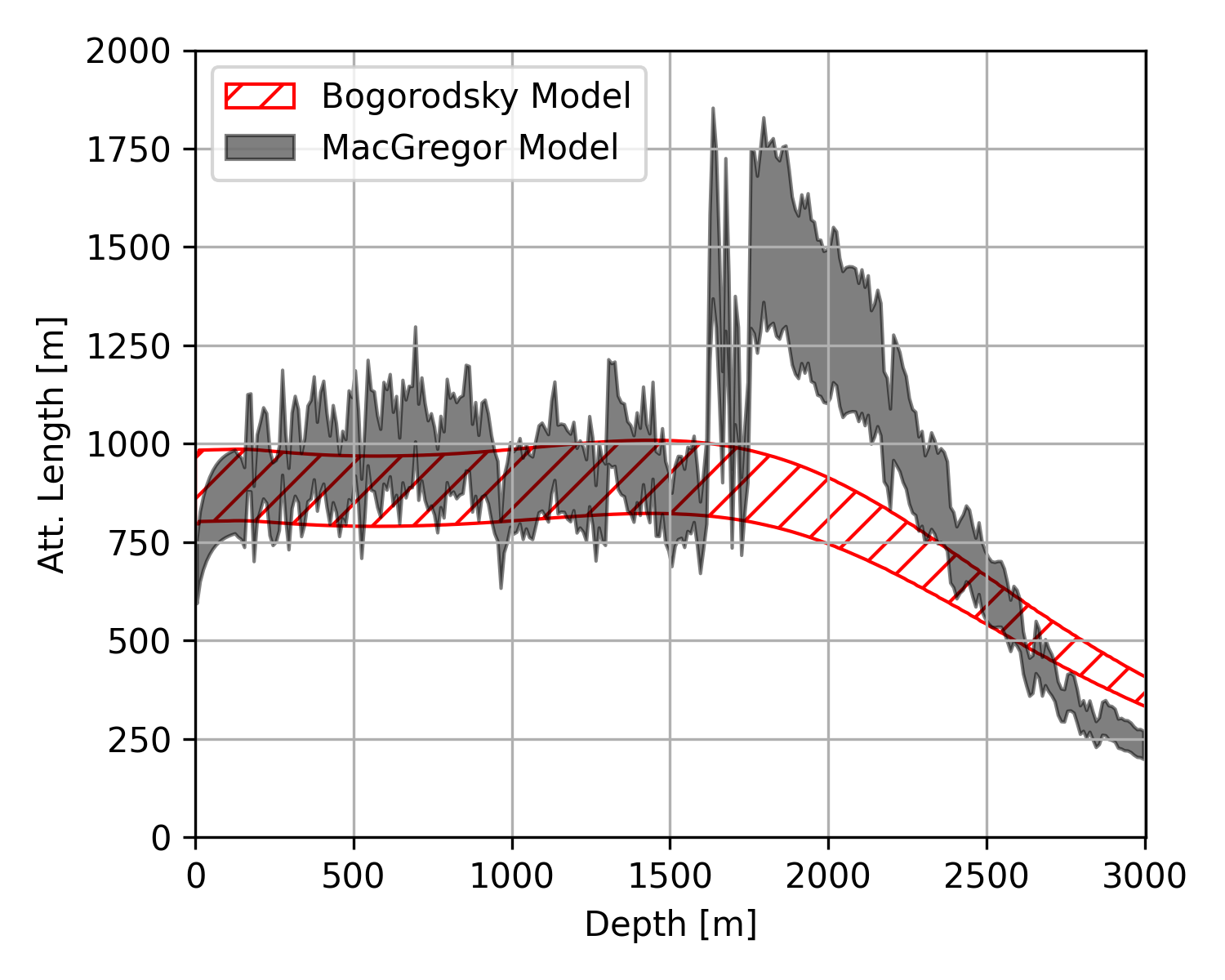}
\caption{Electric field attenuation length as a function of depth at 300~MHz as derived from the two model of ice attenuation, one derived from temperature and chemical impurities \citep{macgregor_2007,macgregor_2015} (black) from temperature alone \citep{Bogorodsky_1985}(red). Hatched and filled regions denotes $\pm1\sigma$. } 
\label{fig:temperature}
\end{figure}

Using the model of electric field attenuation length as a function of depth, we then unfold and solve for the parameter $A(f)$, requiring that the depth-integrated attenuation matches our measured depth-averaged attenuation. The result reported at 300~MHz is plotted in Fig. \ref{fig:temperature}. We note a conspicuous enhancement in the $L_\alpha$(z) profile at depth $\gtrsim$1600 m; this feature tracks a similarly precipitous drop in the tabulated GRIP $H^+$ and $NH_4^+$ molar concentrations at z$\sim$1600 m, resulting in a corresponding enhanced radio-frequency transparency at those depths. 

\cite{Avva:2014ena} derived the electric field attenuation length as a function of depth using a simplified model of the field attenuation dependence on temperature alone. They assume a linear relationship between the log of the attenuation versus the temperature of the ice:

\begin{equation}
    L_{\alpha}(f, T(z)) = A(f) \cdot 10^{m T(z)}.
\end{equation}

The parameter $m$ is taken to be the average of the two sites measured by \cite{Bogorodsky_1985} and set equal to $-0.017$ ($^\circ C$)$^{-1}$. The parameter $A(f)$ is derived from measured value of bulk field attenuation at each frequency $f$ and set equal to 200--280~m. For comparison, our result using the simplified model as reported at 300~MHz is plotted in Fig. \ref{fig:temperature}.

\begin{figure}
\centering
\includegraphics[width=0.48\textwidth]{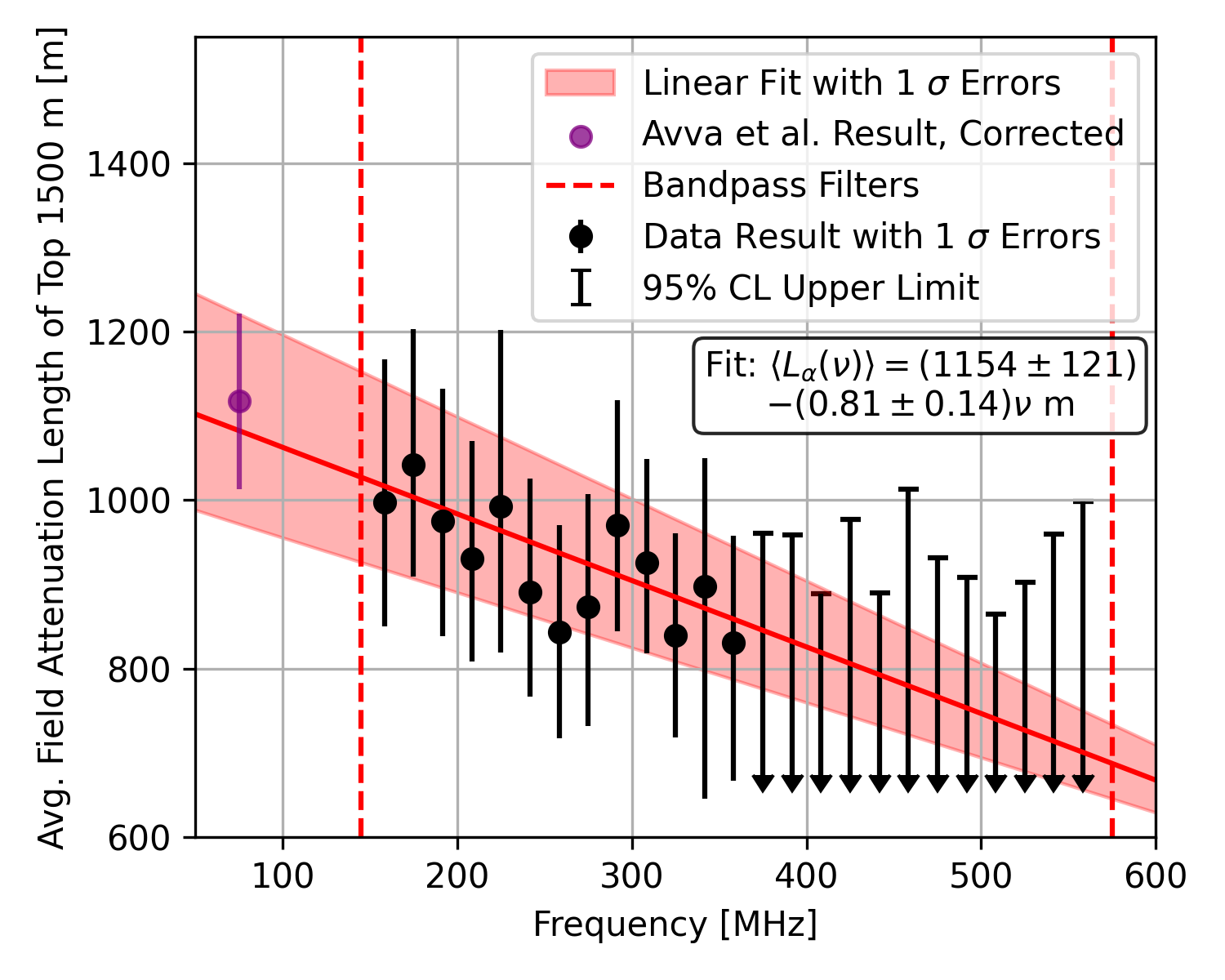}
\caption{Measurement of the average electric field attenuation for the top 1500~m of the ice sheet, as a function of frequency at Summit Station, derived from the measured bulk field attenuation in Fig. \ref{fig:result} and the relationship between attenuation and temperature. Overlaid is the $\pm1\sigma$ confidence interval of a linear fit of the data. Parameters of the fit are described in the text. Frequency $\nu$ in the fit is in units of MHz.}
\label{fig:top_ice}
\end{figure}

The average electric field attenuation length of the top 1500~m of ice is of particular interest to RNO-G as the majority of neutrino interactions detectable by the experiment occur in this region \citep{christoph_energy}. It can be extracted from the bulk result using the field attenuation versus depth relation defined by \cite{macgregor_2007, macgregor_2015}.
The average field attenuation length for the top 1500~m of ice result is shown in Fig. \ref{fig:top_ice}. For reference, the bulk attenuation measurement at 300~MHz is $756^{+71}_{-87}$~m ($-11.49^{+1.49}_{-0.99}$~dB~/~km) while the average attenuation measurement of the top 1500~m of ice is $926^{+107}_{-124}$~m ($-9.38^{+1.45}_{-0.97}$~dB~/~km).

Our measurement can be used in simulations that calculate RNO-G's sensitivity to astrophysical neutrinos. For those simulations, we include, for convenience, a linear fit to the average electric field attenuation length for the top 1500 m of ice, shown in Fig. \ref{fig:top_ice}. 
We find a significant correlation between the two parameters of the linear fit, which yields a slope of $-0.81~\pm~0.14$ m / MHz ($-0.12~\pm~0.02$ dB / km / MHz), intercept of $1154~\pm~121$~m ($-7.53~\pm~0.72$ dB / km), and a correlation coefficient $\rho = -0.95$. 

We compare our obtained result of the average electric field attenuation length for the top 1500 m at 300 MHz of $926^{+107}_{-124}$~m to other similar measurements. \cite{Avva:2014ena} extrapolated their results at 75~MHz to 300~MHz and estimated $\langle L_\alpha \rangle = 1022^{+230}_{-253}$~m, consistent (within uncertainty) with the result presented herein. The electric field attenuation length at the South Pole has been measured, with the focusing factor included, to be $\langle L_\alpha \rangle = 1660^{+255}_{-120}$~m at 300~MHz for the top 1500~m of ice \citep{ALLISON2012457}, a factor 1.8 times longer than our measurement. A linear fit to the log of attenuation length versus temperature (only) from \cite{Bogorodsky_1985}, $20^\circ$C colder ice yields an expected attenuation length ${\sim}2.2$ times longer at South Pole than at Summit, however, the higher concentration of Cl$^-$ at South Pole ($\sim4\mu$M \citep{macgregor_2007} compared to $\sim0.5\mu$M at Summit Station in the top 1500~m of ice) moderates this expectation.

Measurements of the bulk radio field attenuation length at Summit Station have also been performed using air-borne radio sounding data. \cite{macgregor_2015} inferred the bulk radio field attenuation by comparing the relative strengths of internal reflectors, obtaining attenuation lengths of 750--850~m at 150--200~MHz around Summit Station, consistent (within uncertainty) with the result presented herein. \cite{Stockham2016RadioFI} measured the bulk radio field attenuation using the relative strength of the radio echo from the snow surface and the bedrock, obtaining at an attenuation length of $546 \pm 23$~m at 150-200~MHz around Summit Station. This measurement is notably lower than our result and the measurement done by \cite{macgregor_2015} potentially due to radiometric calibration issues \citep{macgregor_2015,Stockham2016RadioFI}. 

The RNO-G experiment, currently under construction at Summit Station in Greenland, is set to be one of the world's largest particle detectors. Our measurement of the bulk electric field attenuation length at Summit Station is consistent with previous measurements, with reduced systematic uncertainties. Our measurement will ultimately increase the precision of RNO-G's UHEN sensitivity estimates, which will either better motivate upper limits in the case of a null result or decrease uncertainties on the measured flux of ultra-high energy neutrinos in the universe in the case of observation.

\subsection{Acknowledgements}
We would like to thank the staff at Summit Station and Polar Field Services for logistical support in every way possible and to our colleagues at the British Antarctic Survey for unending enthusiasm while building and operating the BigRAID borehole drill for RNO-G.

We would like to acknowledge our home institutions and funding agencies for supporting the RNO-G work; in particular the Belgian Funds for Scientific Research (FRS-FNRS and FWO) and the FWO programme for International Research Infrastructure (IRI), the National Science Foundation through the NSF Awards 2118315 and 2112352 and the IceCube EPSCoR Initiative (Award ID 2019597), the German research foundation (DFG, Grant NE 2031/2-1), the Helmholtz Association (Initiative and Networking Fund, W2/W3 Program), the University of Chicago Research Computing Center, and the European Research Council under the European Unions Horizon 2020 research and innovation programme (grant agreement No 805486).

\bibliography{igsrefs}   
\bibliographystyle{igs}  

\end{document}